\documentclass{PoS}
\usepackage{url}

\title{Testing fundamental principles with high-energy cosmic rays}

\ShortTitle{Fundamental principles and cosmic rays}

\author{\speaker{Luis Gonzalez-Mestres}\\
        LAPP, Universit\'e de Savoie, CNRS/IN2P3\\ 
        B.P. 110, 74941 Annecy-le-Vieux Cedex, France\\
        E-mail: \email{lgm\_sci@yahoo.fr}}

\abstract{It is not yet clear \cite{Auger1} whether the observed flux suppression for ultra-high energy cosmic rays (UHECR) at energies above $\simeq $ 4.$10^{19}$ eV is a signature of the Greisen-Zatsepin-Kuzmin (GZK) cutoff or corresponds, for instance, to the maximum energies available at the relevant sources. Both phenomena can be sensitive to violations of standard special relativity modifying cosmic-ray propagation or acceleration at very high energy \cite{Gonzalez-Mestres1,Gonzalez-Mestres2}, and would in principle allow to set bounds on Lorentz symmetry violation (LSV) parameters in models incorporating a privileged local reference frame (the "vacuum rest frame", VRF \cite{Gonzalez-Mestres3,Gonzalez-Mestres4}). But the precise phenomenological analysis of the experimental data is far from trivial, and other effects can be present. The effective parameters can be directly linked to Planck-scale physics, or even to physics beyond Planck scale, as well as to the dynamics and effective symmetries relating LSV mechanisms for nucleons, quarks, leptons and the photon \cite{Gonzalez-Mestres5,Gonzalez-Mestres6}. If a VRF exists, LSV can modify the internal structure of particles at very high energy \cite{Gonzalez-Mestres6,Gonzalez-Mestres7}. Conventional symmetries may also cease to be valid at energies close to the Planck scale. Other possible violations of fundamental principles and conventional basic hypotheses (quantum mechanics, quark confinement, energy and momentum conservation, vacuum homogeneity and "static" properties, effective space dimensions...) can also be considered \cite{Gonzalez-Mestres7,Gonzalez-Mestres8} and possibly tested in high-energy cosmic-ray experiments. Even below UHE (ultra-high energy), exotic signatures cannot be excluded \cite{Gonzalez-Mestres9,Gonzalez-Mestres10}. We present an updated discussion of the theoretical and phenomenological situation, including prospects for earth-based and space experiments and a simple potential interpretation of the observed UHECR composition in terms of LSV where the GZK cutoff would be replaced by spontaneous emission of photons or e$^+$ e$^-$ pairs. As the OPERA result \cite{OPERA} on a possible superluminal propagation of the muon neutrino was announced after the conference, we briefly comment on the consistency problems \cite{Gonzalez-Mestres11,Gonzalez-Mestres12} that a $\simeq $ 2.5 x $10^{-5}$ critical speed anomaly for the muon neutrino can raise taking into account well-established experimental evidence and astrophysical observations.}

\FullConference{The 2011 Europhysics Conference on High Energy Physics-HEP 2011,\\
		July 21-27, 2011\\
		Grenoble, Rh\^one-Alpes, France}

\begin{document}
\section{Ultra-high energy cosmic rays, Planck scale and new physics}

The current standard relativity principle was formulated by Henri Poincar\'e \cite{Poincar\'e} on the grounds of the result of the Michelson-Morley experiment. It replaced the Galilean principle by a new one where the speed of light $c$ is a universal critical speed and the Maxwell equations are covariant. Electromagnetism has turned out to be a fundamental interaction of conventional elementary particles, and Poincar\'e's relativity remains valid today. But as already pointed out by Albert Einstein as early as 1921 \cite{Einstein}, such a standard relativity may be broken beyond some very small distance scale. A crucial question is that of the existence of a VRF. Contrary to the old ether, the vacuum of quantum field theory is a real medium where fields can condense through mechanisms initially inspired by solid state physics. For small wave vectors, condensed matter physics naturally generates Lorentz-like symmetries with critical speeds of the order of that of sound \cite{Gonzalez-Mestres3,Gonzalez-Mestres4}. They are deformed at larger frequencies, and the privileged rest frame manifests itself through this deformation. In our Universe, the hypothesis of the existence of a VRF appears at least as legitimate as a "perfect" space-time symmetry obtained by modifying the Poincar\'e algebra of special relativity.             

The experimental properties of UHECR can contain observable effects generated at the Planck scale or beyond it \cite{Gonzalez-Mestres4}, and allowing for a study of the new physics associated to such scales. An illustrative example can be obtained taking for UHE particles, with LSV and a VRF, the quadratically deformed dispersion relation \cite{Gonzalez-Mestres1,Gonzalez-Mestres4}: 
\begin{equation}
E ~ \simeq ~ p~c~+~m^2~c^3~(2~p)^{-1}~-~p~c~\alpha ~(p~ c ~E_a^{-1})^2/2
\end{equation}
\noindent
where $E$ is the energy, $p$ the momentum, $c$ the speed of light, $m$ the mass, $\alpha $ a constant standing for the deformation strength and $E_a$ the fundamental energy scale at which the deformation is generated. Then, the (negative) deformation term $\Delta ~E~\simeq ~-~p~c~\alpha ~(p~ ~c ~E_a^{-1})^2/2$ becomes larger than the (positive) mass term $m^2~c^3~(2~p)^{-1}$ above a transition energy $E_{trans}$ given by \cite{Gonzalez-Mestres1,Gonzalez-Mestres6}:
\begin{equation}
E_{trans} ~ \simeq ~ \alpha ^{-1/4}~(E_a~m)^{1/2}~c
\end{equation}
For a proton, with $\alpha $ = 1, $E_{trans} $ = $10^{20}$ eV for $E_a ~\simeq ~ 10^3 ~ E_{Planck}$ where $E_{Planck}$ is the Planck energy. For an electron, the same effect would occur at $E_{trans} $ = $10^{19}$ eV if $E_a ~\simeq $ 2 x $10^4 ~ E_{Planck}$. A 10$^{19}$ eV muon would similarly be sensitive to physics generated at $E_a ~\simeq ~ 10^2 ~ E_{Planck}$. Above $E_{trans}$, we expect the effective internal structure and interaction properties of particles to be modified \cite{Gonzalez-Mestres6,Gonzalez-Mestres7}. 

In traditional grand-unification patterns, standard particle symmetries are expected to become more exact as the energy scale increases and masses become comparatively smaller. However, in the patterns considered here, this property would hold only below some critical scale like $E_{trans}$. Above $\approx ~ E_{trans}$, new physics can manifest itself including possible signatures of an ultimate (superbradyonic ?) composite structure \cite{Gonzalez-Mestres4,Gonzalez-Mestres7}. New properties of conventional particles, as well as hidden differences between them, may become apparent and invalidate lower-energy symmetries. 

The possibility that new physics generated at the Planck scale or at a deeper fundamental scale becomes apparent above $\approx ~ E_{trans}$ provides a major motivation for a long-term experimental effort, ground based and through satellites, to study the highest-energy cosmic rays as well as their potential sources. All usually admitted fundamental principles of Physics and properties of standard particles should be tested as far as possible, including energy and momentum conservation, quantum mechanics, vacuum properties and stability, particle propagation in vacuum... UHECR data and statistics are not yet enough for this purpose, but they already provide important information. 

\section{Examples of possible signatures}

With a VRF close to that naturally suggested by cosmic microwave background radiation (CMB), the rest frames used in UHECR and other experiments will move at nonrelativistic speed with respect to the VRF. Then, an equation like (1.1) remains stable under standard Lorentz transformations between such frames, up to corrections much smaller than the deformation term. Such transformations naturally preserve the additivity and conservation of energy and momentum. 

As pointed out in \cite{Gonzalez-Mestres3,Gonzalez-Mestres4}, patterns with a VRF allow for phenomena that would be forbidden otherwise. This is the case for energy-dependent spontaneous decays. Particles that are stable at low energy can become unstable at higher energy, and conversely. Similarly, a typical signature of new physics would be the suppression of the GZK cutoff due to LSV with a VRF, as first proposed in \cite{Gonzalez-Mestres1,Gonzalez-Mestres4}. Such an effect would be directly related to the $E_{trans}$ transition and to the fact that the deformation $\Delta ~E$ becomes larger than the energy of the relevant CMB photons. 

However, if the proton is to be dealt with like a composite object, his deformed kinematics can have a value of $\alpha $ smaller than those of quarks and gluons, similarly to those of nuclei as compared to the proton \cite{Gonzalez-Mestres6,Gonzalez-Mestres7}. Then, the possible existence of the GZK cutoff would still be compatible with a significant LSV at the Planck scale. Also, if the UHE proton has a smaller value of $\alpha $ than the photon, it can spontaneously decay by emitting a photon if the photon $\Delta ~E$ becomes larger than the proton mass term. If the photon has a lower $\alpha $ than the electron, it can decay into an e$^+$ e$^-$ pair. 

The same LSV pattern would simultaneously lead to a suppression of synchrotron radiation at the highest-energy sources provided that energies $\approx ~ E_{trans}$ can be reached \cite{Gonzalez-Mestres2}. Higher energies than conventionally expected can then be produced at the astrophysical sources. 

Present data \cite{Auger1} show a fall of the UHECR spectrum compatible with the GZK cutoff, but the effect can also be explained by other mechanisms. A LSV pattern of the type (1.1) inhibiting the GZK cutoff but allowing for spontaneous emission of gamma rays or e$^+$ e$^-$ pairs by hadrons and nuclei can potentially account for the observed composition at the highest energies, where heavier nuclei appear to dominate. If the deformation of the photon or electron energy is to be compared with the mass term of the emitting proton or nucleus, the observed mass composition of the UHE Auger spectrum can naturally emerge, as heavier nuclei will start decaying at higher energies.

The consistency of the present approach also requires paying attention to lower energies. The OPERA collaboration reports \cite{OPERA} that muon neutrinos between a few GeV and more than 100 GeV appear to travel at a speed $c ~ (1 ~+ ~\delta )$ where $\delta ~= ~ (2.48 ~\pm ~0.28)$ x 10$^{-5}$. Such a strong LSV at comparatively low energy does not seem to fit with the pattern considered here. The OPERA neutrino would present at least two kinds of consistency problems \cite{Gonzalez-Mestres11} : i) spontaneous decays, emitting in particular electron-positron pairs ; ii) to be able to decay by emitting the superluminal neutrino, the charged pion should present a similar critical speed anomaly that would propagate to hadrons and cosmic rays. These points have been further developed in \cite{Gonzalez-Mestres12,CohenGlashow}. Taking the pion critical speed to be equal to $c$, a trivial bound can be immediately obtained \cite{Gonzalez-Mestres12} before introducing lepton and neutrino masses:
\begin{equation}
p_{\nu } ^2~\leq ~m_{\pi} ^2~c ^3~[2~\delta ~c]^{-1}
\end{equation}
($ p_{\nu } $ = neutrino momentum, $m_{\pi} $ = pion mass). One then gets $p_{\nu } ~\leq ~ 20~GeV/c$, in contradiction with the OPERA data. $\delta ~= ~ 10^{-6}$ would yield $ p_{\nu }~\leq $ 100 GeV. Bounds can be made stronger using the muon mass \cite{Gonzalez-Mestres12}. Consistency with SN 1987A data is also a problem for the OPERA result.

\section{Conclusion}

As already emphasized in \cite{Gonzalez-Mestres1,Gonzalez-Mestres4}, UHECR experiments are a powerful microscope directly focused on the Planck scale and beyond. Long-term experimental programs and permanent observatories are required, allowing for better statistics with more precise analyses and theoretical studies. All basic principles of standard particle physics should be tested in this way.

\end{document}